\documentclass[11pt]{article}
\usepackage[utf8]{inputenc}
\usepackage[T1]{fontenc}
\usepackage{amsmath,amssymb}
\usepackage{graphicx}
\usepackage{booktabs}
\usepackage{siunitx}
\usepackage{array}
\usepackage{adjustbox}
\usepackage{authblk}
\usepackage[margin=1in]{geometry}
\usepackage{xcolor}
\usepackage{caption}
\usepackage[numbers,sort&compress]{natbib}
\usepackage[colorlinks=true,linkcolor=blue,citecolor=blue,urlcolor=blue]{hyperref}

\title{\textbf{A Reproducible Pipeline for Symmetry-Respecting Excited States
on Near-Term Quantum Computers: The H$_2$O/STO-3G Case}}

\author{Huajing Song\thanks{\texttt{wilsonsong85@gmail.com} \protect\newline
\texttt{Huajing.Song@prattwhitney.com}}}
\affil{Materials \& Process Engineering Division, Pratt \& Whitney,
An RTX Business, East Hartford, CT, USA}
\date{\today}

\begin{document}
\maketitle

\begin{abstract}
Variational excited-state quantum algorithms fail for reasons usually studied in
isolation: barren plateaus, symmetry contamination, finite-sampling instability,
and hardware cost. Using one small but complete system---H$_2$O in the STO-3G
basis (12 qubits, Jordan--Wigner)---we assemble these into a single reproducible
pipeline, checking every claim against exact diagonalization. The bare qubit
Hamiltonian interleaves cation ($N{=}7$) states below the neutral manifold;
hardware-efficient and number-conserving ans\"atze stall at Hartree--Fock, an
exact stationary point by Brillouin's theorem, while ADAPT-VQE escapes;
variational deflation inherits the contamination and inverts the spectrum, whereas
the quantum equation-of-motion (qEOM) subspace method restores the ladder to
sub-milli-Hartree accuracy. Particle number is protected \emph{structurally} under
shot noise, and a realistic measurement model collapses the thousands of subspace
matrix elements to $\sim\!10^5$ commuting groups; a matrix-aware shot allocation
then reaches chemical accuracy at $\sim\!3\times10^9$ total shots---a thousandfold
below the naive per-element estimate and reachable in days---leaving single-circuit
gate fidelity, not measurement, as the binding constraint. This work is a teaching
and benchmarking reference, not a new method; all code, parameters, and figures are
released.
\end{abstract}

\section{Introduction}
\label{sec:intro}

The calculation of molecular excited states is a natural target for quantum
computers and a recurring testbed for variational quantum algorithms
\citep{mcardle2020,peruzzo2014}. Unlike the ground state, excited states demand
that an algorithm resolve states that are close in energy, often of different
spin or charge character, and do so on hardware whose dominant error sources are
finite sampling and two-qubit gate infidelity. A substantial literature now
documents the individual obstacles---barren plateaus and trainability
\citep{mcclean2018}, symmetry contamination of variational ans\"atze
\citep{gard2020,greenediniz2021}, the statistical fragility of generalized
eigenvalue problems \citep{ollitrault2020,gevp2026}, and the daunting
measurement cost of expectation-value algorithms. These obstacles are usually
demonstrated one at a time, on different systems, which makes it hard for a
newcomer to see how they fit together for a single molecule.

This paper is deliberately not a new method. Its contribution is a single,
end-to-end, fully reproducible pipeline for one well-chosen
system---H$_2$O in the minimal STO-3G basis---that exhibits \emph{every} one of
these pathologies and their resolution, with each numerical claim checked
against an exact reference and released as runnable code. We chose H$_2$O/STO-3G
precisely because it is small enough to diagonalize exactly (so every quantum
result has a ground truth) yet rich enough to display particle-number
contamination, spin contamination, ansatz trainability failure, deflation
failure, and the full measurement-cost question.

We state our scope plainly. The mechanisms below are known: the shot-noise
behavior of qEOM was analyzed for this exact system by Ollitrault et al.\
\citep{ollitrault2020}; the conditioning of the noisy generalized eigenvalue
problem was analyzed recently in Ref.~\citep{gevp2026}; charge-sector mixing in
qubit Hamiltonians is standard in the tapering literature
\citep{bravyi2017}; and a noise-robust successor that removes the overlap matrix
entirely (q-sc-EOM) already exists \citep{asthana2022}. Our value is
integration, verification, and reproducibility---and the honest display of
failure modes that are often omitted. Section~\ref{sec:system} fixes the system
and reference data; Section~\ref{sec:spectral} diagnoses the spectral
contamination; Section~\ref{sec:ground} treats ground-state trainability;
Section~\ref{sec:excited} contrasts excited-state methods;
Section~\ref{sec:measurement} develops the measurement strategy and shot
allocation; Section~\ref{sec:sampling} treats finite sampling and overlap
conditioning; Section~\ref{sec:hardware} gives the resource estimate; and
Section~\ref{sec:conclusions} concludes.

\section{System, mapping, and reference data}
\label{sec:system}

We study a single water molecule at its equilibrium geometry in the STO-3G
minimal basis. Freezing the oxygen $1s$ core leaves 8 active electrons (4
spin-up, 4 spin-down) in 6 spatial orbitals, i.e.\ 12 spin-orbitals, which we map
to 12 qubits with the Jordan--Wigner transformation. The resulting qubit
Hamiltonian is a weighted sum of Pauli strings,
\begin{equation}
\hat H = \sum_{k=1}^{551} c_k\,\hat P_k,
\qquad \hat P_k \in \{I,X,Y,Z\}^{\otimes 12},
\label{eq:ham}
\end{equation}
whose $\ell_1$ coefficient norm,
\begin{equation}
\lVert H\rVert_1 = \sum_k \lvert c_k\rvert = \SI{46.36}{\hartree},
\label{eq:l1}
\end{equation}
sets the finite-sampling cost in Section~\ref{sec:sampling}.

Three symmetry operators play a central role: the total particle number
$\hat N$, the spin projection $\hat S_z=\tfrac12(\hat N_\alpha-\hat N_\beta)$,
and the total spin $\hat S^2$. Physical neutral states have $\langle\hat
N\rangle=8$; singlets, doublets, and triplets have $\langle\hat
S^2\rangle=0,\,0.75,\,2$ respectively. Our reference values, computed classically
with PySCF \citep{pyscf2018} and by exact diagonalization, are an exact
(CASCI) ground-state energy of \SI{-75.009047}{\hartree} and a Hartree--Fock
energy of \SI{-74.961}{\hartree}; the constant nuclear-repulsion-plus-core shift
is \SI{-51.398662}{\hartree}. All electronic-structure objects (Hamiltonian,
mapper, symmetry operators) are built by a shared module and every script emits a
provenance manifest recording package versions and parameters
(Appendix~\ref{app:repro}).

\section{Spectral diagnosis: charge-sector contamination}
\label{sec:spectral}

The first and most instructive observation requires no quantum algorithm at all.
Diagonalizing the bare 12-qubit Hamiltonian, the low-lying spectrum does
\emph{not} consist of the neutral excited states one might expect. Several of the
lowest excited eigenstates have $\langle\hat N\rangle=7$ and $\langle\hat
S^2\rangle=0.75$---they are doublet states of the H$_2$O$^+$ \emph{cation},
which sit \emph{below} the neutral ($N{=}8$) excited manifold. This is not a
bug: the qubit Hamiltonian acts on the full $2^{12}$-dimensional Fock space,
which contains every particle-number sector, and nothing in $\hat H$ forbids a
cation state from lying low in energy. The same fact is well known in the qubit
tapering literature, where the lowest eigenvalue of a molecular Hamiltonian may
belong to a charged species \citep{bravyi2017}.

\begin{figure}[t]
\centering
\includegraphics[width=0.7\linewidth]{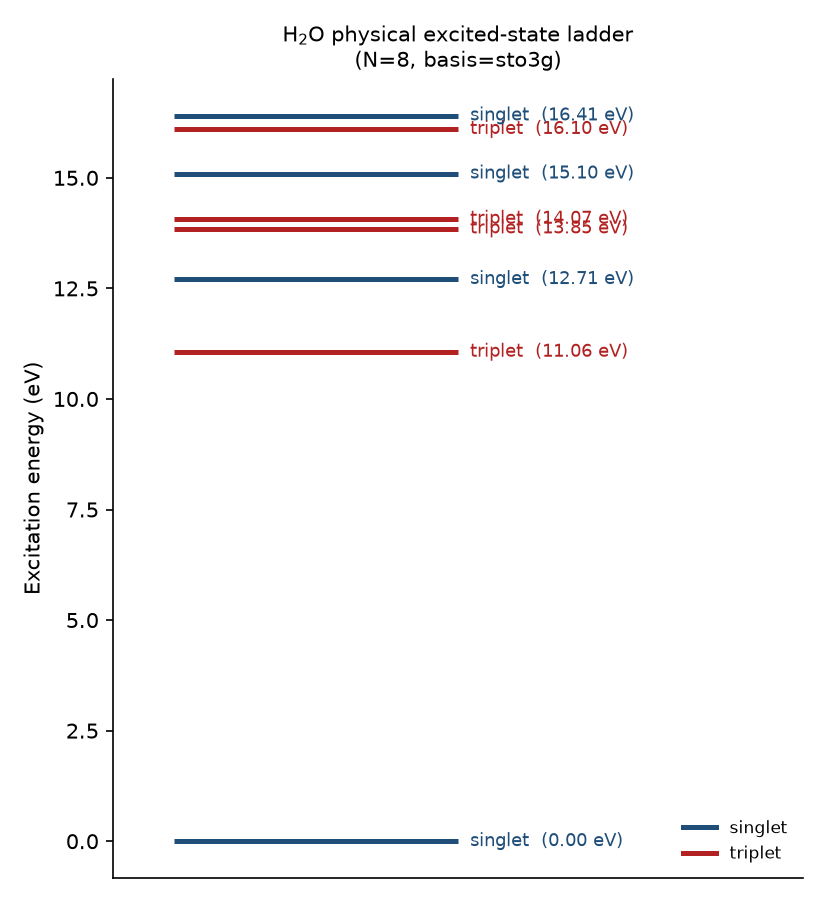}
\caption{Physical neutral excited-state ladder of H$_2$O/STO-3G recovered by
symmetry-penalty projection onto the $N{=}8$, $S_z{=}0$ sector, with spin labels.
The bare (unprojected) spectrum interleaves H$_2$O$^+$ cation doublets
($\langle\hat N\rangle=7$) below these states.}
\label{fig:ladder}
\end{figure}

To recover the physical neutral ladder we project onto the $N{=}8$, $S_z{=}0$
sector by diagonalizing the penalized Hamiltonian
\begin{equation}
\hat H_\mu = \hat H + \mu\big(\hat N-8\big)^2 + \mu\,\hat S_z^2 .
\label{eq:proj}
\end{equation}
The resulting ladder (Fig.~\ref{fig:ladder}) is the ground truth
against which every excited-state method below is measured: a triplet at
\SI{11.06}{\electronvolt}, the first optically accessible singlet at
\SI{12.71}{\electronvolt}, two triplets at \SI{13.85}{} and
\SI{14.07}{\electronvolt}, and a singlet at \SI{15.10}{\electronvolt}. The
practical lesson is immediate: any method that searches the full Hilbert space
without enforcing $\hat N$ can---and, as we show next, does---collapse onto these
spurious cation states.

\section{Ground state: trainability of three ans\"atze}
\label{sec:ground}

\begin{figure}[t]
\centering
\includegraphics[width=0.9\linewidth]{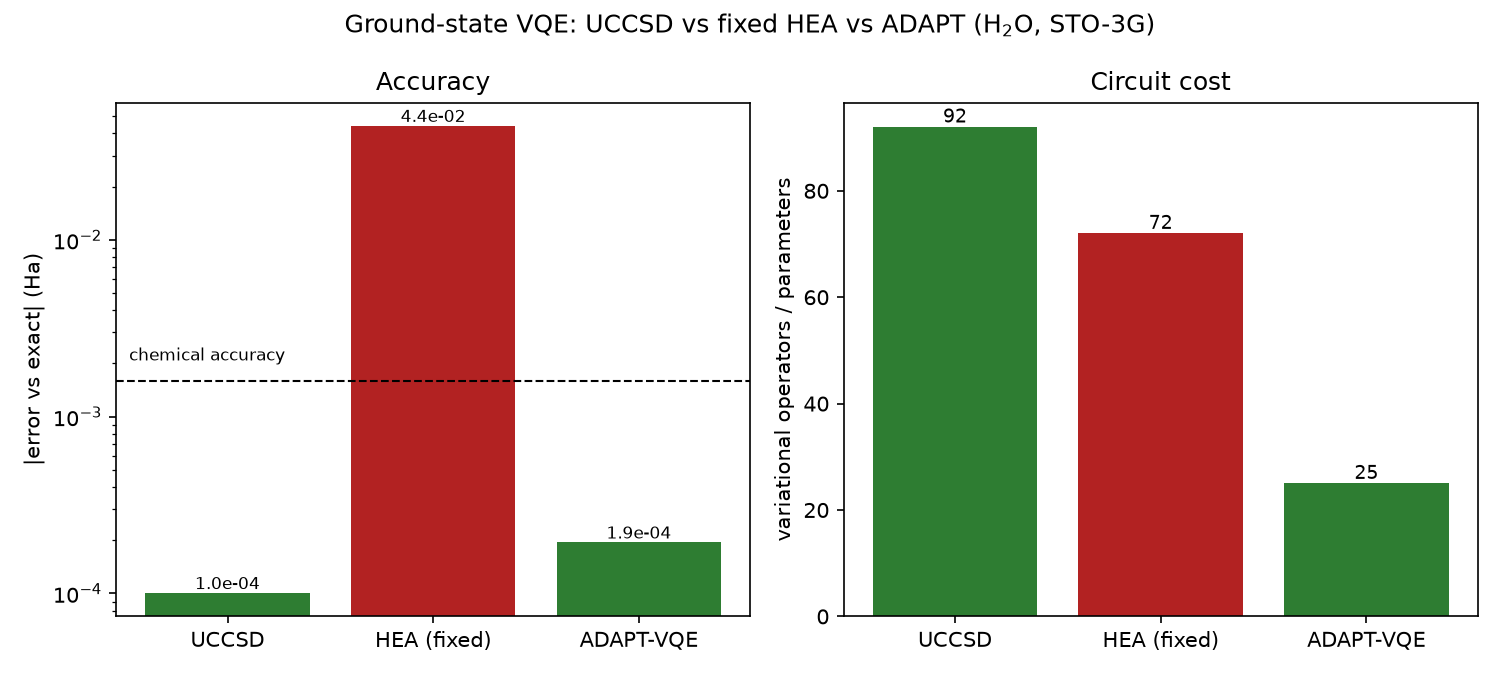}
\caption{Ground-state convergence for three ans\"atze. UCCSD and ADAPT-VQE reach
chemical accuracy (shaded); hardware-efficient and number-conserving ans\"atze
stall at Hartree--Fock with vanishing gradient.}
\label{fig:ground}
\end{figure}

Before excited states, we examine how three families of ansatz reach (or fail to
reach) the ground state from a Hartree--Fock start (Fig.~\ref{fig:ground}).

A chemically motivated unitary coupled-cluster ansatz (UCCSD, 92 variational
parameters) reaches near-chemical accuracy, converging to
\SI{-75.008946}{\hartree} (error $\sim\!10^{-4}$~Ha) with $\langle\hat N\rangle=8$
and $\langle\hat S^2\rangle=0$ preserved throughout.

By contrast, two hardware-friendly ans\"atze fail in an identical and
diagnostic way. A generic hardware-efficient ansatz (EfficientSU2, 72
parameters) and a number-conserving ansatz (ExcitationPreserving) both stall
exactly at the Hartree--Fock energy, with zero energy variance across many random
initializations. The cause is structural, not a tuning problem: the energy
gradient at the Hartree--Fock state is \emph{exactly zero} for these circuits,
for both linear and full entanglement maps. This is Brillouin's theorem---single
excitations do not couple to the Hartree--Fock determinant through the
Hamiltonian, and these ans\"atze, built from one- and two-qubit rotations, fail
to inject the double excitations that do. The Hartree--Fock state is therefore an
exact stationary point, and gradient-based optimization cannot leave it. This is
Brillouin's theorem,
\begin{equation}
\langle\mathrm{HF}|\hat H|\Phi_i^a\rangle = 0,
\label{eq:brillouin}
\end{equation}
which makes the single-excitation gradient vanish at the reference
(Appendix~\ref{app:brillouin}); only doubles carry nonzero gradient there.

The fix is to make double excitations available from the start. ADAPT-VQE
\citep{grimsley2019} grows the ansatz
\begin{equation}
|\psi(\boldsymbol\theta)\rangle
  = \prod_{k} e^{i\theta_k \hat A_k}\,|\mathrm{HF}\rangle,
\qquad \hat A_k = i\big(\hat T_k - \hat T_k^\dagger\big),
\label{eq:adapt}
\end{equation}
adding at each step the pool operator of largest energy gradient
\begin{equation}
g_k = \left.\frac{\partial E}{\partial\theta_k}\right|_{\theta_k=0}
  = \langle\psi|[\hat H,\hat A_k]|\psi\rangle
  = 2\,\mathrm{Im}\,\langle\psi|\hat H\,\hat T_k|\psi\rangle .
\label{eq:adaptgrad}
\end{equation}
It selects a double excitation as its very first operator and immediately descends
below Hartree--Fock. With 25 selected operators (24 doubles and a single single
excitation) it reaches \SI{-75.008853}{\hartree} (error $\sim\!1.9\times10^{-4}$~Ha)
with clean symmetry. The dominance of doubles is itself the fingerprint of
Brillouin's theorem: singles have vanishing gradient at Hartree--Fock and are
not selected until the state has already moved away from it.

\section{Excited states: failure versus fix}
\label{sec:excited}

\subsection{Variational deflation on a hardware-efficient ansatz fails}
Variational quantum deflation (VQD) \citep{higgott2019} finds excited states by
augmenting the energy with overlap penalties against previously found states,
\begin{equation}
F_k(\boldsymbol\theta)
  = \langle\psi_k|\hat H|\psi_k\rangle
  + \sum_{i<k}\beta_i\,\big|\langle\psi_i|\psi_k\rangle\big|^2 .
\label{eq:vqd}
\end{equation}
Applied with a hardware-efficient ansatz, it fails in four compounding ways: it
recovers the wrong ``ground'' state (the Hartree--Fock energy, per
Section~\ref{sec:ground}); states one and two converge onto the spurious
$N{=}7$ cation doublets of Section~\ref{sec:spectral}; the genuine neutral
excited states are missed entirely; and the resulting ordering is inverted
relative to the physical ladder. As a control, the deflation overlaps themselves
are numerically sound (down to $10^{-10}$), confirming that the failure is not
the deflation machinery but the unconstrained ansatz exploring the contaminated
full Hilbert space. That HEA tends to collapse onto lower-lying states of the
wrong symmetry has been noted before \citep{greenediniz2021}; here it is made
quantitative and tied directly to the cation contamination.

\subsection{Symmetry penalties are a diagnostic, not a clean cure}
A natural repair is to add symmetry penalties to the deflation cost. We find this
is genuinely instructive but not a clean fix. A particle-number penalty removes
the cation states (all states return to $\langle\hat N\rangle=8$); but $\hat
S_z=0$ alone does not fix spin, leaving singlet--triplet mixtures
($\langle\hat S^2\rangle\approx1$). Adding a strong (quartic) $\hat S^2$ penalty
does enforce the target spin, but degrades trainability and energies---an
aggressive spin penalty trades symmetry purity against optimizability. The
honest conclusion is that penalty methods expose the contamination clearly but do
not robustly resolve it; in our hands the strong-$\hat S^2$ configuration that
restored the correct spin also corrupted the energies, so we treat the penalty
study as a diagnostic rather than a fix.

\subsection{qEOM restores the correct ordering}
\begin{figure}[t]
\centering
\includegraphics[width=0.95\linewidth]{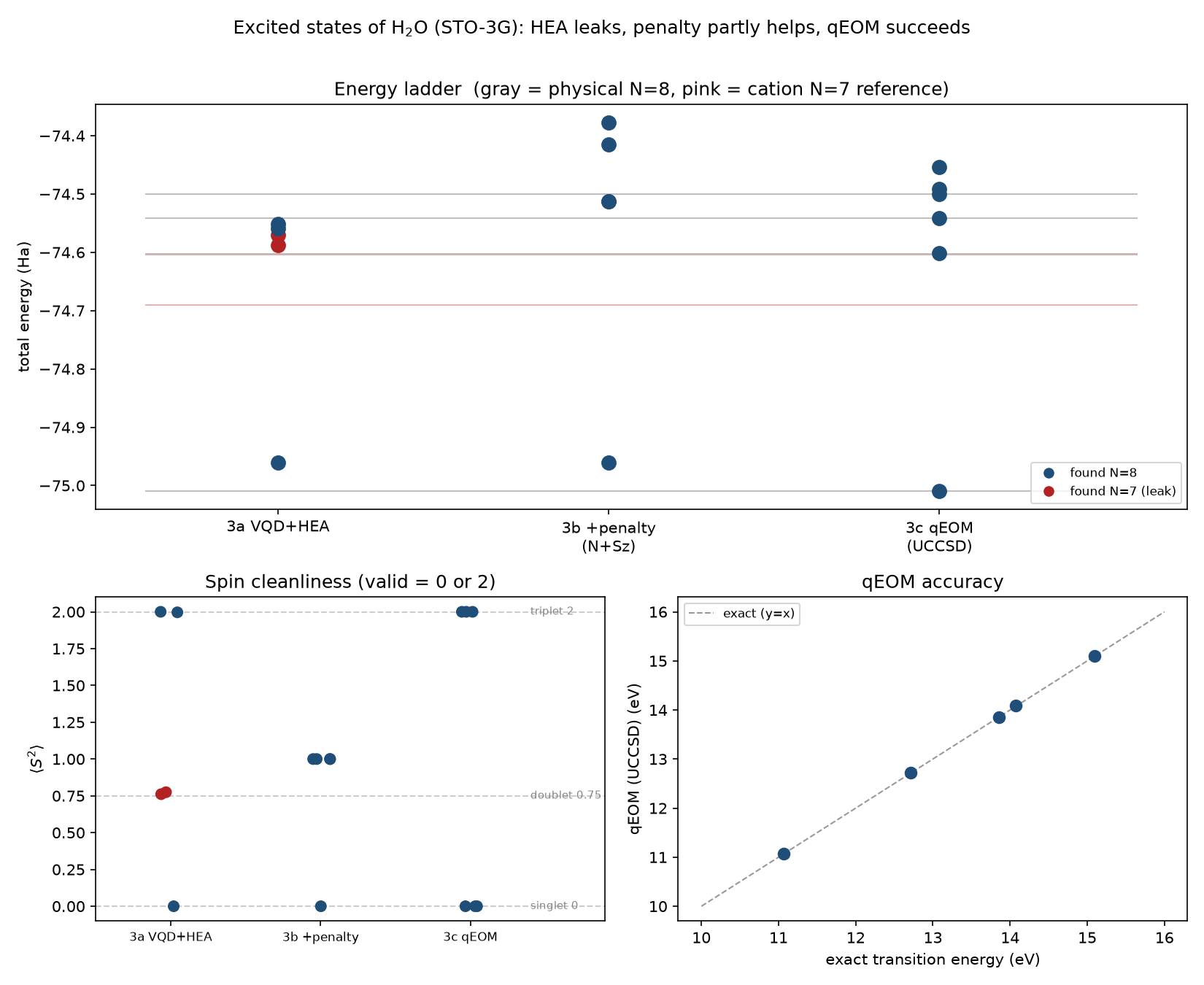}
\caption{Excited-state methods. Left to right: VQD on a hardware-efficient ansatz
yields cation-contaminated, mis-ordered states; qEOM, built from
number-conserving excitation operators, reproduces the exact neutral ladder.
Supplementary comparisons (symmetry-penalty tradeoff; exact-vs-UCCSD-ground
robustness) are in Appendix~\ref{app:qeom}, Fig.~\ref{fig:excited_supp}.}
\label{fig:excited}
\end{figure}

The robust resolution is to work in a symmetry-respecting subspace. The quantum
equation-of-motion (qEOM) method \citep{ollitrault2020}, in its subspace-expansion
form (quantum subspace expansion, QSE) \citep{mcclean2017}, builds a basis $\{|\psi_0\rangle,\,\hat
E_\mu|\psi_0\rangle\}$ from excitation operators applied to the ground state and
solves the generalized eigenvalue problem
\begin{equation}
H\,\mathbf c = E\,S\,\mathbf c,
\label{eq:gevp}
\end{equation}
with projected Hamiltonian and overlap matrices
\begin{equation}
H_{\mu\nu}=\langle\psi_0|\hat E_\mu^\dagger \hat H \hat E_\nu|\psi_0\rangle,
\qquad
S_{\mu\nu}=\langle\psi_0|\hat E_\mu^\dagger \hat E_\nu|\psi_0\rangle .
\label{eq:hsmatrix}
\end{equation}
Because the excitation operators conserve particle number by construction, the
entire basis lives in the $N{=}8$ sector and cation contamination is structurally
impossible. The method reproduces the exact ladder to sub-milli-Hartree accuracy
for both an exact and a UCCSD ground state (Fig.~\ref{fig:excited}), and is
robust to ground-state error: a UCCSD ground state (error $\sim\!10^{-4}$~Ha)
yields gaps within a few meV of those from the exact ground state.

\section{Measurement strategy: Pauli reuse, commuting groups, and shot allocation}
\label{sec:measurement}

Before asking what shot noise does to the spectrum, we must ask what it actually
costs to measure the subspace at all. The qEOM/QSE working equations require every
entry of the projected Hamiltonian $H$ and overlap $S$ matrices, each an
expectation value on the prepared ground state. For the eight-root ADAPT-QSE
problem the subspace has dimension 93, giving $4{,}371$ unique upper-triangle
entries in each of $H$ and $S$. Each entry expands into Pauli expectations on the
checkpointed ADAPT state,
\begin{equation}
H_{\mu\nu} = \sum_k c_k^{(\mu\nu)}\,\langle\psi_0|\hat P_k|\psi_0\rangle,
\label{eq:paulielem}
\end{equation}
and across all entries these reference $2{,}121{,}464$ distinct Pauli strings
$\hat P_k$---a number that, taken at face value, is the origin of the
``measurement wall'' folklore.

\paragraph{The structure collapses the cost.}
That face value is misleading, because the same Pauli expectation appears in many
matrix entries. Deduplicating globally, then sorting the unique strings into
qubit-wise-commuting (QWC) bases that can be read simultaneously, reduces
$2.1\times10^6$ strings to $126{,}469$ measurement bases; the $551$-term
Hamiltonian alone collapses to $313$ groups. We verify the decomposition by
reconstructing the clean $H$ and $S$ matrices from the grouped Pauli values,
recovering them to $\sim\!3\times10^{-15}$. Global Pauli reuse and commuting-group
measurement are standard measurement-compression tools
\citep{ikhtiarudin2025}; what matters here is that for a real
excited-state subspace they convert a nominally $10^6$-fold measurement problem
into a $10^5$-fold one before a single shot is allocated.

\begin{figure}[t]
\centering
\includegraphics[width=0.9\linewidth]{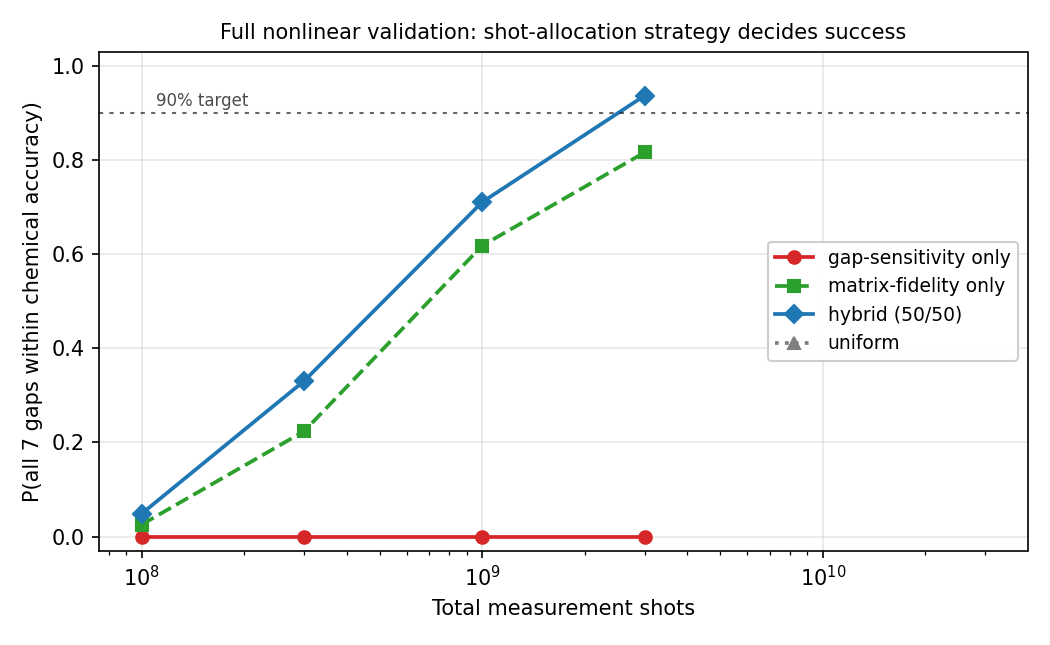}
\caption{Shot-allocation strategy in the full nonlinear ADAPT-QSE solve. The
gap-sensitivity-only policy that screens well under linear response fails
completely (success probability pinned at zero); a matrix-aware hybrid is the
robust winner, reaching $94\%$ all-gaps-within-chemical-accuracy at
$\sim\!3\times10^9$ total shots. Uniform allocation reaches only $\sim\!60\%$ at
$3\times10^{10}$ shots (single marker, far right).}
\label{fig:measurement}
\end{figure}

\paragraph{Allocation is decisive---and the obvious heuristic fails.}
Given a fixed total shot budget, how should shots be distributed across the
$126{,}469$ groups? We compare three policies under a minimum per-group floor:
\emph{uniform}; \emph{gap-sensitivity}, weighting each group by its first-order
influence on the seven excitation gaps; and \emph{matrix-aware} (and hybrids
thereof), weighting by each group's contribution to the variance of the $H$ and
$S$ entries themselves. Each policy assigns group $g$ a shot count
\begin{equation}
n_g = \max\!\left(n_{\rm floor},\
  \big(N_{\rm tot}-N_{\rm floor}\big)\,
  \frac{w_g}{\sum_{g'} w_{g'}}\right),
\label{eq:alloc}
\end{equation}
differing only in the weight $w_g$: constant for uniform, gap-sensitivity for the
gap policy, matrix-element variance for the matrix-aware policy, or a sum of the
two for the hybrid. In a linear-response screening model the gap-sensitivity
policy looks excellent. It then \emph{fails outright} in the full nonlinear
generalized-eigenvalue solve (Fig.~\ref{fig:measurement}): starving the
gap-insensitive matrix elements destroys the identity of the Ritz roots, so the
solved spectrum is meaningless (maximum root RMSE $\sim\!9.4$~eV and dominant-state
fidelity $\sim\!0.25$ at every budget, with zero successful trials). The lesson is
that what must be measured well is not what the eigenvalues are locally sensitive
to, but what keeps the projected problem solvable. A \emph{hybrid} policy that
splits the budget evenly between matrix-fidelity and gap-sensitivity is the
robust winner, reaching a $94\%$ probability that all seven gaps land within
chemical accuracy ($\SI{0.0435}{\electronvolt}$) at $\sim\!3\times10^9$ total
shots; pure matrix-fidelity allocation reaches $\sim\!82\%$ there, and uniform
allocation manages only $\sim\!60\%$ even at $3\times10^{10}$ shots.

\paragraph{The wall is real but an order of magnitude lower than advertised.}
The headline number is that $\sim\!3\times10^9$ \emph{total} shots---not per
element---suffice for chemical accuracy on the full seven-gap ladder. The naive
accounting that multiplies a per-element requirement by $4{,}371$ elements
overstates the budget by more than three orders of magnitude, because it ignores
the reuse and grouping that the problem's structure provides for free. This does
not abolish the measurement cost, but it relocates the binding constraint: as the
resource analysis below shows, $3\times10^9$ shots is hours-to-days of wall-clock
on the fastest hardware, whereas a single high-fidelity state preparation is not.
The result is consistent with recent hardware-oriented studies of grouped,
variance-allocated measurement for excited-state subspace methods
\citep{sundar2026}.

\section{Finite sampling: structural protection and overlap conditioning}
\label{sec:sampling}

On hardware, the qEOM matrix elements are expectation values estimated from a
finite number of measurement shots. The previous section built the realistic
measurement model---per-Pauli variances, reuse, and commuting groups. Here we
complement it with a deliberately pessimistic \emph{screening} channel that is
useful precisely because it is the worst case: each matrix element is perturbed by
\emph{independent} Gaussian noise,
\begin{equation}
\tilde X_{ij} = X_{ij} + \varepsilon_{ij},
\qquad
\varepsilon_{ij}\sim\mathcal N\!\big(0,\ \sigma_X^2/N_{\rm shots}\big),
\quad X\in\{H,S\},
\label{eq:noise}
\end{equation}
with $\sigma_H=\lVert H\rVert_1=\SI{46.36}{\hartree}$ (the standard $\ell_1$
measurement-cost scale) and $\sigma_S=1$, with no reuse and no correlation. We
Monte-Carlo the generalized eigenvalue solve over many realizations at each shot
count. The contrast between this independent-noise caricature and the structured
model of Section~\ref{sec:measurement} is itself instructive.

\begin{figure}[t]
\centering
\includegraphics[width=0.95\linewidth]{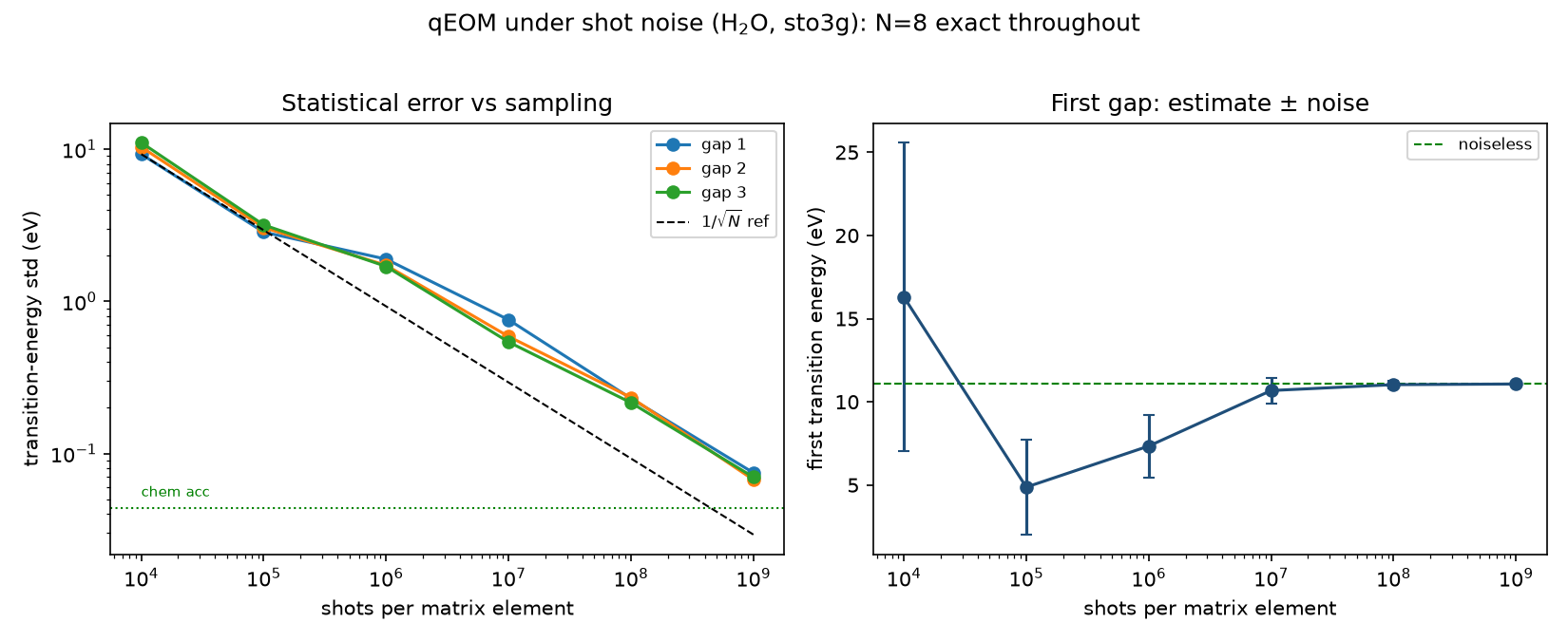}
\caption{qEOM under shot noise. Particle number is exactly $N{=}8$ at all shot
counts (structural protection); transition-energy statistical error falls with
sampling but reaches chemical accuracy only near $10^9$ shots per matrix element.}
\label{fig:shotnoise}
\end{figure}

\paragraph{Particle number is protected structurally.}
The single most robust result of this study is that $\langle\hat N\rangle=8$
\emph{exactly}, in 100\% of realizations, at \emph{every} shot count
(Fig.~\ref{fig:shotnoise}). Because every basis vector $\hat E_\mu|\psi_0\rangle$
conserves $\hat N$, any solution of the generalized eigenvalue problem is a
combination of $N{=}8$ states; sampling noise can blur energies but cannot leak
the result into the wrong particle-number sector. This is the precise sense in
which qEOM's symmetry protection is structural rather than statistical, and it is
the sharpest contrast with VQD, where noise could push the state toward the
cation sector. Spin cleanliness, by contrast, improves gradually with shots.

\begin{figure}[t]
\centering
\includegraphics[width=0.95\linewidth]{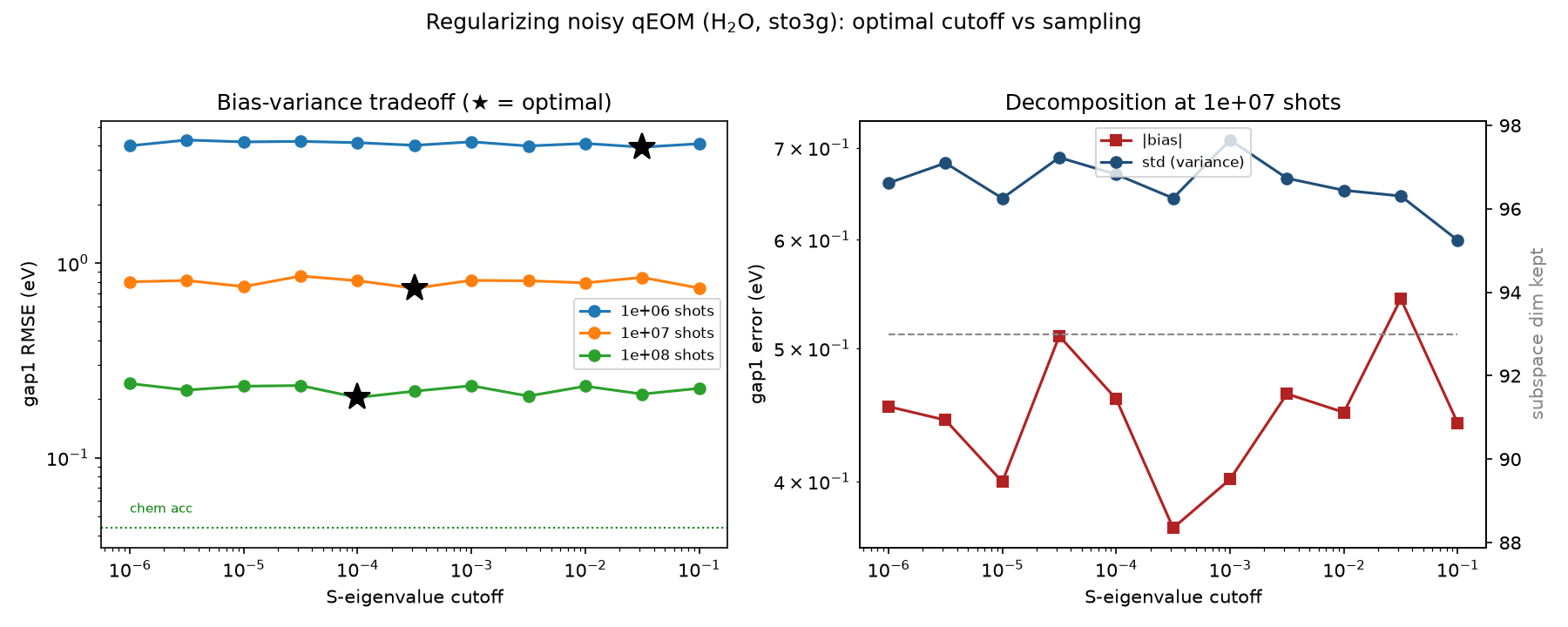}
\caption{Overlap-matrix regularization. Left: gap-1 RMSE versus $S$-eigenvalue
cutoff for several shot counts, with the RMSE-optimal cutoff marked; the optimum
moves toward smaller cutoffs as shots increase. Right: bias--variance
decomposition.}
\label{fig:regularization}
\end{figure}

\paragraph{Regularization helps the worst case, but the realistic matrix is well-conditioned.}
In the independent-noise screening model the transition energies are biased at low
shot counts: noisy small eigenvalues of the overlap matrix $S$ manufacture
spurious low-energy solutions of the generalized eigenvalue problem, the
conditioning effect analyzed in Ref.~\citep{gevp2026}. The standard remedy is
canonical-orthogonalization regularization---discarding $S$-eigenvalue directions
below a cutoff before solving---and sweeping the cutoff
(Fig.~\ref{fig:regularization}) exposes a clean bias--variance tradeoff whose
RMSE-optimal value tracks the noise floor, falling from $\sim\!3\times10^{-2}$ at
$10^6$ shots/element to $\sim\!10^{-4}$ at $10^8$. The important caveat, and a
correction to the impression this screening model gives, is that the realistic
measurement of Section~\ref{sec:measurement} is \emph{far} better behaved: with the
actual per-Pauli covariances and a matrix-aware allocation, the sampled $S$ stays
positive-definite in every trial with condition number $\approx 1.005$, and the
regularization is never triggered. The dramatic ill-conditioning is thus largely
an artifact of the independent-noise caricature; on real grouped measurements it
is a mild effect, not a wall. Regularization with $\varepsilon\sim10^{-6}$ remains
sensible routine practice \citep{gevp2026}, but it is insurance, not the load-bearing step.

\paragraph{Reconciling the two cost estimates.}
The screening model, taken literally, reproduces the folklore measurement wall:
chemical accuracy on the first transition would need $\sim\!10^9$ shots
\emph{per element}, and na\"ively multiplying by $\sim\!4{,}300$ elements gives a
budget of order $10^{13}$. Section~\ref{sec:measurement} shows why that number is
an overestimate---it assumes every element is measured independently, with no
reuse and no grouping. The structured budget that actually applies is
$\sim\!3\times10^9$ \emph{total} shots, more than three orders of magnitude
smaller. The screening model is still worth reporting because it bounds the
damage when correlations are unfavorable and isolates the conditioning mechanism;
but the honest cost of the method is the structured one. For a route that sidesteps
the overlap-conditioning question entirely, q-sc-EOM replaces $S$ by the identity
and is correspondingly noise-free in the overlap \citep{asthana2022}.

\section{Hardware resource estimate}
\label{sec:hardware}

Finally we translate the algorithm into compiled resources on best-in-class 2026
hardware in three modalities, transpiling the ground-state preparation circuits
to each device's native gate set and connectivity. We take, as best-in-class
anchors (accessed 2026-06-23): superconducting, Google Willow, two-qubit fidelity
99.88\%, native CZ, restricted planar lattice \citep{willow2026}; trapped-ion,
Quantinuum Helios, two-qubit fidelity 99.921\%, native $ZZ(\theta)$, all-to-all
connectivity \citep{helios2025}; and neutral-atom, Atom Computing, two-qubit
fidelity 99.6\%, native Rydberg CZ, reconfigurable connectivity
\citep{atomcomputing2026}. Two-qubit gate times span roughly tens of nanoseconds
(superconducting), $\sim\!1\,\mu$s (neutral atom), and tens of microseconds
(trapped-ion).

\begin{figure}[t]
\centering
\includegraphics[width=0.98\linewidth]{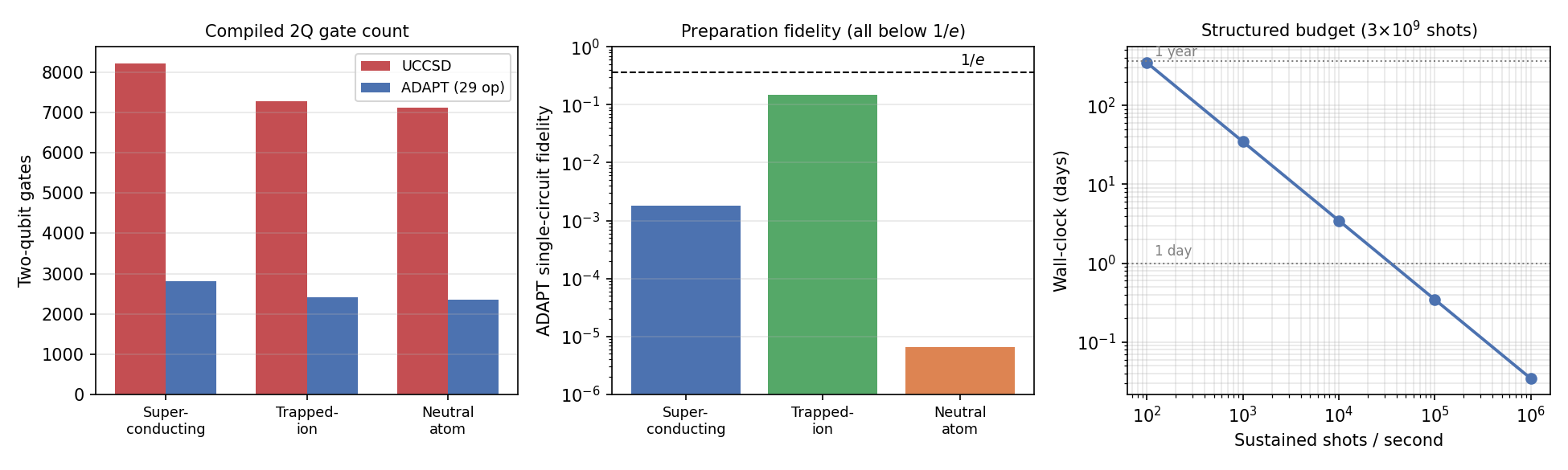}
\caption{Hardware resource estimate from the actual 29-operator ADAPT checkpoint.
Left: compiled two-qubit gate counts, UCCSD versus ADAPT, by modality. Middle:
single-circuit fidelity of ADAPT preparation (all below $1/e$). Right: wall-clock
to complete the structured $\sim\!3\times10^9$-shot measurement budget versus
sustained execution rate; days at $10^4$ shots/s, with one-day and one-year
references. The measurement budget is reachable; the per-circuit fidelity is the
binding constraint.}
\label{fig:resource}
\end{figure}

Table~\ref{tab:resource} and Fig.~\ref{fig:resource} summarize the result. These
figures use the actual 29-operator ADAPT checkpoint that drives the excited-state
and measurement analysis; the ground-state demonstration of
Section~\ref{sec:ground} reached chemical accuracy already at 25 operators, and the
slightly longer checkpoint is the one carried downstream. ADAPT's compactness
matters: it requires roughly a third as many two-qubit gates as UCCSD
($\sim\!2{,}400$ versus $\sim\!7{,}500$). UCCSD ground-state preparation is
hopeless on any current device (single-circuit fidelity $\le10^{-3}$). Even ADAPT
exceeds every device's fidelity budget---its single-circuit fidelity is
$\sim\!15\%$ on trapped-ion, $\sim\!0.2\%$ on superconducting, and
$\sim\!7\times10^{-6}$ on neutral atom---so even the compact ansatz needs error
mitigation to yield a usable ground state. Trapped-ion fares best here because it
combines the highest fidelity with all-to-all connectivity (zero routing overhead;
the superconducting heavy-hex lattice adds several hundred routing-induced
two-qubit gates).

The wall-clock picture is where the structured measurement analysis changes the
conclusion. The naive accounting---$\sim\!10^9$ shots/element against
$\sim\!4{,}300$ elements, executed serially---gives the dispiriting figures of
tens to tens-of-thousands of years that earlier discussions of this method
report. But Section~\ref{sec:measurement} showed the real budget is
$\sim\!3\times10^9$ \emph{total} shots, and one shot is one state preparation and
measurement. At a sustained execution rate of $10^4$ shots/s---well within reach
of superconducting hardware---that budget completes in $\sim\!3.5$ days; at
$10^3$/s, $\sim\!35$ days; finishing within a calendar year requires only
$\sim\!95$ shots/s. The measurement budget, in other words, is not the wall. What
remains binding is single-circuit fidelity: every one of those $3\times10^9$
preparations is an ADAPT circuit whose noiseless fidelity is already well below
$1/e$ on all three modalities (Table~\ref{tab:resource}), so each shot returns a
heavily corrupted state unless error mitigation intervenes. The honest verdict is
therefore the inverse of the folklore: the symmetry-protected method is correct,
its overlap matrix is well-conditioned under realistic measurement, and its shot
budget is reachable in days---but its per-circuit gate error is not, and that is
what motivates error mitigation today and fault tolerance tomorrow (vendor-stated
targets: IBM by 2029, Quantinuum by the end of the decade, cited as targets
rather than endorsed predictions).

\begin{table}[t]
\centering
\caption{Compiled ground-state-preparation resources (12 qubits) for the actual
29-operator ADAPT checkpoint and the UCCSD/HEA comparators. Two-qubit and
single-qubit gate counts after transpilation (optimization level 3) to each
native gate set and connectivity; the ``routing'' column is the two-qubit
overhead the restricted superconducting (heavy-hex) lattice adds over all-to-all;
single-circuit fidelity $=f_{2Q}^{n_{2Q}}f_{1Q}^{n_{1Q}}$. Reproduces on
qiskit 1.4.6.}
\label{tab:resource}
\small
\begin{tabular}{llrrrr}
\toprule
Ansatz & Modality & 2Q gates & Routing 2Q & 1Q gates & Single-circ.\ fid. \\
\midrule
UCCSD & Superconducting & 8226 & 1155 & 31581 & $3.9\times10^{-9}$ \\
UCCSD & Trapped-ion     & 7279 & 0    & 24152 & $3.0\times10^{-3}$ \\
UCCSD & Neutral atom    & 7119 & 0    & 7751  & $1.7\times10^{-16}$ \\
ADAPT & Superconducting & 2806 & 467  & 9787  & $1.8\times10^{-3}$ \\
ADAPT & Trapped-ion     & 2418 & 0    & 7383  & $1.5\times10^{-1}$ \\
ADAPT & Neutral atom    & 2357 & 0    & 2486  & $6.6\times10^{-6}$ \\
\bottomrule
\end{tabular}
\end{table}

\section{Conclusions}
\label{sec:conclusions}

For a single, exactly solvable molecule we have traced the full arc of
variational excited-state quantum simulation: a bare Hamiltonian spectrum
contaminated by cation states; hardware-efficient ans\"atze frozen at
Hartree--Fock by Brillouin's theorem; deflation inheriting the contamination and
inverting the spectrum; symmetry penalties diagnosing but not curing the problem;
a symmetry-respecting subspace method (qEOM) restoring the exact ladder; particle
number protected structurally under shot noise while energies remain
statistics-limited; a realistic measurement model in which Pauli reuse and
commuting groups, paired with a matrix-aware shot allocation, bring chemical
accuracy on all seven gaps to $\sim\!3\times10^9$ total shots---over a
thousandfold below the naive estimate, and reachable in days---while leaving
single-circuit gate fidelity as the binding obstacle. Along the way a plausible
allocation heuristic (pure gap sensitivity) was shown to fail in the full
nonlinear solve, a cautionary result we report rather than bury. None of these
mechanisms is individually new; the contribution is to show them together, in one
reproducible pipeline, with the failures displayed rather than hidden, and to
correct the over-pessimistic measurement-wall accounting that a per-element
caricature invites. We hope it serves as a teaching reference and a benchmark, and
we point readers toward the noise-robust reformulations (q-sc-EOM and
measurement-reduction techniques) and error-mitigated hardware that the analysis
identifies as the path forward.

\section*{Acknowledgments}
This work was supported by internal Strategic Initiative funding and computing
resources provided by P\&W, RTX.

\section*{Data and code availability}
\begin{sloppypar}
All code, parameters, and figure-regeneration scripts are released at
\url{https://github.com/WilsonSong360/H2O-VQE-GS_EX}. Each script emits a
provenance manifest recording package versions and run parameters; key pinned
dependencies are listed in Appendix~\ref{app:repro}.
\end{sloppypar}

\appendix
\section{Reproducibility}
\label{app:repro}
The pipeline is organized as shared modules (system construction, provenance I/O)
plus per-step scripts, each writing JSON, CSV, Markdown, figures, and a manifest.
The exact software stack used to produce every number and figure in this paper,
as recorded in the run manifests, is:

\begin{center}
\small
\begin{tabular}{ll}
\toprule
Component & Version \\
\midrule
Python        & 3.12.10 \\
qiskit        & 1.4.6 \\
qiskit-nature & 0.7.2 \\
qiskit-algorithms & 0.3.1 \\
PySCF         & 2.13.1 \\
NumPy / SciPy & 2.5.0 / 1.18.0 \\
Matplotlib    & 3.11.0 \\
\bottomrule
\end{tabular}
\end{center}

\begin{sloppypar}
Each step script is run from its directory against the shared modules (e.g.\
\texttt{python step2d\_adapt\_vqe.py}, \texttt{python step5\_shot\_noise.py},
\texttt{python resource\_estimate.py}); figures regenerate into the per-step
\texttt{results/} folders alongside their manifests.
\end{sloppypar}

\section{Brillouin stationarity of the Hartree--Fock state}
\label{app:brillouin}
For an adaptive ansatz with Hermitian generators $\hat A=i(\hat T-\hat
T^\dagger)$, the energy gradient at a reference $|\psi\rangle$ is $2\,\mathrm{Im}
\langle\psi|\hat H\hat A|\psi\rangle$. At the Hartree--Fock determinant the
single-excitation contribution vanishes by Brillouin's theorem, so circuits that
supply only single-like rotations have a vanishing gradient and the
Hartree--Fock state is an exact stationary point; only double excitations carry
nonzero gradient there. Consistent with this, the ADAPT-VQE run selects a double
excitation at its first iteration and reaches an error of
$1.9\times10^{-4}$~Ha relative to the exact ground state after 25 operators
(24 doubles and a single single excitation), with $\langle\hat N\rangle=8$ and
$\langle\hat S^2\rangle\approx2\times10^{-4}$; the lone single excitation is not
selected until iteration~23, after the state has already left the Hartree--Fock
point.

\section{qEOM subspace construction and the generalized eigenvalue solve}
\label{app:qeom}
\begin{figure}[h]
\centering
\includegraphics[width=0.95\linewidth]{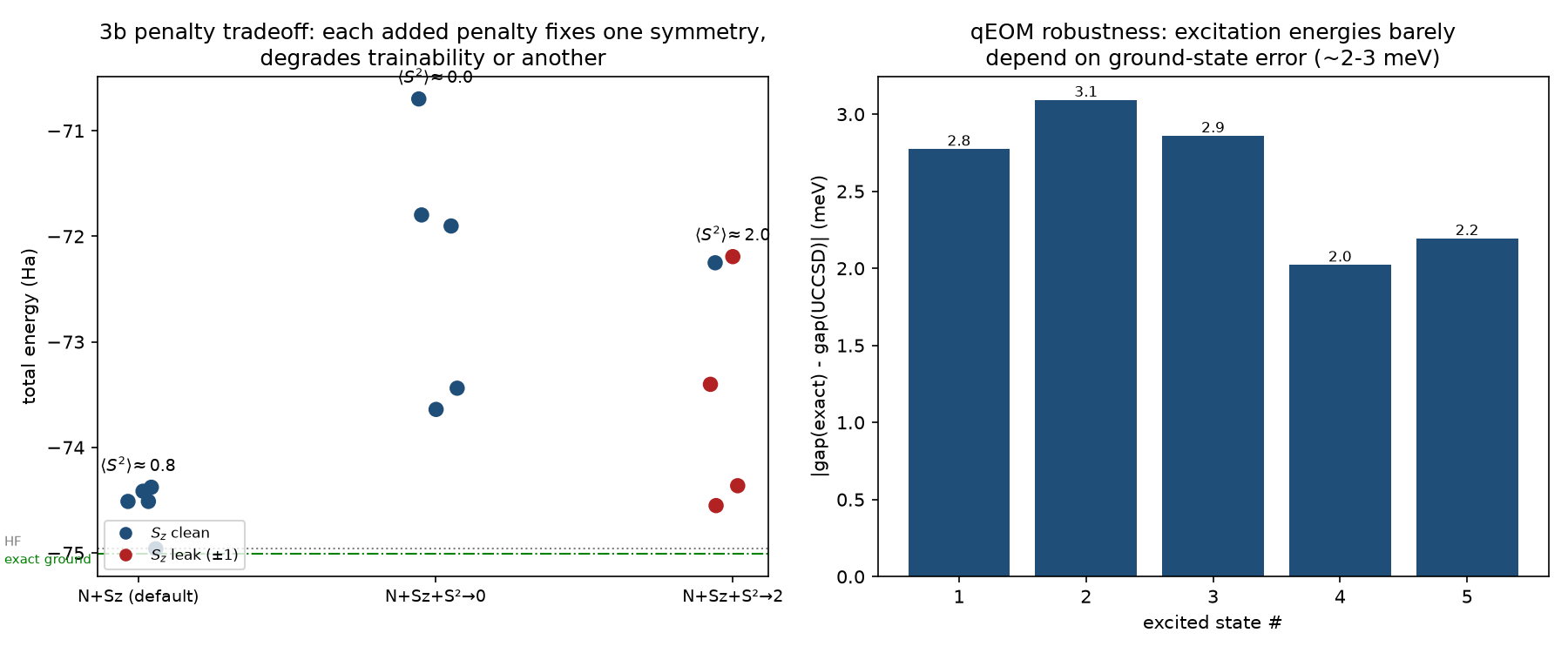}
\caption{Supplementary excited-state comparisons: penalty-method tradeoff across
configurations, and qEOM robustness to ground-state error (exact versus UCCSD
ground).}
\label{fig:excited_supp}
\end{figure}

We use the subspace-expansion form with bare single and double excitation
operators, which conserve particle number by construction. The generalized
eigenvalue problem is solved by canonical orthogonalization with an
$S$-eigenvalue cutoff. Figure~\ref{fig:excited_supp} shows the symmetry-penalty
tradeoff and the exact-versus-UCCSD-ground robustness.

\section{Shot-noise and regularization model}
\label{app:noise}
Matrix elements are perturbed by independent symmetric Gaussian noise of standard
deviation $\sigma/\sqrt{N_{\rm shots}}$ ($\sigma_H=\lVert H\rVert_1$,
$\sigma_S=1$); the generalized eigenvalue problem is solved per realization and
aggregated over Monte-Carlo trials. The shot-noise study uses $n_{\rm mc}=200$
realizations over the shot grid $\{10^4,10^5,\dots,10^9\}$ shots per matrix
element. The regularization sweep uses $n_{\rm mc}=200$ over an eleven-point
logarithmic cutoff grid from $10^{-6}$ to $10^{-1}$ at shot counts
$\{10^6,10^7,10^8\}$, reporting the bias, variance, and RMSE of the first
transition energy against the noiseless qEOM reference ($11.08$~eV; the
exact-diagonalization value is $11.06$~eV, the sub-mHa difference being the
subspace-truncation error of qEOM).

\section{Realistic measurement model and shot allocation}
\label{app:measurement}
The eight-root ADAPT-QSE subspace has dimension 93, giving $4{,}371$
upper-triangle entries in each of the projected $H$ and $S$ matrices. Each entry
is expanded into Pauli expectations on the checkpointed 29-operator ADAPT state;
deduplicating globally yields $2{,}121{,}464$ distinct Pauli strings, sorted into
$126{,}469$ qubit-wise-commuting measurement bases (the $551$-term Hamiltonian
alone gives $313$ groups). The clean $H$ and $S$ matrices reconstruct from the
grouped values to $\sim\!3\times10^{-15}$, a strict correctness check.

Under a fixed total shot budget with a per-group floor, three allocation policies
are compared: \emph{uniform}; \emph{gap-sensitivity}, weighting groups by the
first-order response of the seven gaps; and \emph{matrix-aware} hybrids, weighting
by each group's contribution to the variance of the matrix entries. Validation
uses the full nonlinear generalized-eigenvalue solve over $2\times10^4$
Monte-Carlo realizations per budget, scoring the probability that all seven gaps
fall within chemical accuracy ($\SI{0.0435}{\electronvolt}$). The
gap-sensitivity-only policy, which screens well in a linear-response surrogate,
fails outright here (maximum root RMSE $\sim\!9.4$~eV, dominant-state fidelity
$\sim\!0.25$, zero successful trials at every budget): it starves the matrix
elements that preserve Ritz-root identity. The even-split hybrid reaches $94\%$
success at $\sim\!3\times10^9$ total shots; matrix-fidelity-only reaches
$\sim\!82\%$ and uniform $\sim\!60\%$ (the latter only at $3\times10^{10}$). With
the matrix-aware allocation the sampled overlap matrix is positive-definite in
every trial with condition number $\approx\!1.005$ and the canonical-orthogonalization
cutoff is never triggered. Wall-clock follows from throughput: at a sustained
$10^4$ shots/s the $3\times10^9$-shot budget completes in $\sim\!3.5$ days, and
reaching it within a year requires only $\sim\!95$ shots/s.

\bibliographystyle{unsrtnat}
\bibliography{references}

\end{document}